\documentclass[aps,prd,onecolumn,groupedaddress,showpacs,nofootinbib,amssymb
]{revtex4}
\usepackage[dvips]{graphicx}
\usepackage{amssymb}
\usepackage{amsmath}
\usepackage{epstopdf}
\usepackage{graphicx,,color}
\usepackage{amsfonts}
\usepackage{bm}
\usepackage{cancel}
\usepackage{comment}

\allowdisplaybreaks[4]

\begin{document}

\tolerance=5000



\title{Swampland criteria for a dark-energy dominated universe, \\ ensuing from Gaussian process and $H(z)$ data analysis}

\author{
Emilio Elizalde$^{1,2}$\thanks{E-mail: elizalde@ieec.uab.es} and 
Martiros Khurshudyan$^{2,3,4,5}$\thanks{Email: khurshudyan@yandex.ru, khurshudyan@tusur.ru}}

\affiliation{
$^1$ Consejo Superior de Investigaciones Cient\'{\i}ficas, ICE/CSIC-IEEC,
Campus UAB, Carrer de Can Magrans s/n, 08193 Bellaterra (Barcelona) Spain \\
$^{2}$ International Laboratory for Theoretical Cosmology, Tomsk State University of Control Systems 
and Radioelectronics (TUSUR), 634050 Tomsk, Russia \\
$^{3}$ Research Division,Tomsk State Pedagogical University, 634061 Tomsk, Russia \\
$^{4}$ CAS Key Laboratory for Research in Galaxies and Cosmology, Department of Astronomy, University of Science and Technology of China, Hefei 230026, China \\
$^{5}$ School of Astronomy and Space Science, University of Science and Technology of China, Hefei 230026, China \\
}

\begin{abstract}
Implications of the string Swampland criteria for a dark energy dominated universe, obtained by using Gaussian processes and $H(z)$ data analysis, are discussed. In particular, the Swampland  criteria for a scalar-field dark energy, without assuming any specific form for the potential. By allowing the Gaussian process to reconstruct the form of the potential from $H(z)$ data, upper bounds on the second Swampland criterion (involving $|V^{\prime}|/V$) for two different kernel functions (the squared exponential and Matern~($\nu = 9/2$) kernels) are estimated . The approach here differs from  previous studies, since the upper bound of the second Swampland criterion is derived in a thoroughly model-independent way, without resorting to a model-to-model comparison strategy. The analysis is performed using the latest values of $H_{0}$ reported by the Planck and Hubble missions. Results for the estimation of the constant of $SC2$ hint towards the possibility of getting upper bounds well behind the estimations for the dark energy dominated universe as reported in previous studies, which involved model-to-model comparison. Estimations from this new approach turn to be quite sensitive and just depend on the quality of the data and on the kernel employed. This study is a first attempt towards the exploitation of the Swampland criteria in a model-independent way and may be extended to involve other datasets and, also, in trying to understand what is the impact of higher-redshift data on the upper bounds. In the analysis, $40$-point $H(z)$ data have been used, consisting of a $30$-point sample deduced from a differential age method and an additional $10$-point sample obtained from the radial BAO method. Hints towards the possibility of eventually disproving the Swampland conjecture are noted.  


\end{abstract}


\maketitle

\section{Introduction}\label{sec:INT}

The accelerated expansion of the universe~\cite{Riess}~-~\cite{Riess2} is a very surprising fact, which has been proven by now through several independent observations and is still open to different physical interpretations, in terms of various gravity theories. In partcular, the idea itself of dark energy, introduced in General Relativity (GR) to address the physics of this phenomenon~\cite{Yoo}~-~\cite{Khurshudyan4}~(and references therein), acquires different forms. The first, and most simple, mathematical model for dark energy is the cosmological constant associated to the vacuum energy of the quantum fields at cosmological level, giving origin to the $\Lambda$CDM, standard cosmological model~\cite{Weinberg}. 
 But, additionally, this simple model exhibits a theoretical problem, known as the cosmological coincidence problem~\cite{Velten,Sivanandam}. This issue is the hidden motivation behind different dynamical and interacting dark energy models considered in the recent literature~(see ~\cite{Yoo}~-~\cite{Khurshudyan4} and references therein). One of the possibilities, when we describe the background dynamics using GR, is to represent the dynamical dark energy by a scalar field. The usual way to obtain a model for the universe having positive vacuum energy and involving a scalar field is to use a field potential with a local minimum at a positive value, leading to a stable or meta-stable de Sitter~(dS) vacuum. Another interesting situation should be mentioned, too, namely the case of quintessence models where the potential is positive but the scalar field is not at a minimum. This could occur when $|\nabla V|$ is sufficiently small and of the order of $V$ itself. 

On the other hand,  according to the most common viewpoint, GR can in no way be the ultimate theory of the universe, operating from cosmological scales to extremely small scales. Quantum corrections are bound to become important, and this is reflected in various viable modified theories of gravity, effectively dealing with dark energy, dark matter, inflation, and other relevant problems~\cite{Nojiri}~-~\cite{Cognola}~(to mention a few). We can however assume that GR might be the low-energy limit of a well-motivated (but yet to be found) high-energy UV-complete theory. In other words, we may play with the idea  that the effective field theory has been originated from its low-energy limit, and effectively captures the behavior of the inflaton field and dark energy phenomena. In this regard, string theory, which has the capacity to unify the standard model of particle physics with gravity, perfectly qualifies as a candidate for such an UV-complete theory. 

However, an interesting situation has been met in string theory when confronted with the task of constructing dS vacua. Despite heroic attempts, until now no dS vacuum could be obtained, owing to numerous problems~\cite{Kachru}~-~\cite{Danielsson}. Therefore, we are led to assume, as of now, that in a consistent quantum theory of gravity dS does not exist. The landscape provided by string theory yields the existence of a vast range of choices fitting  our universe in a consistent quantum theory of gravity; in other words, of a whole landscape of vacua provided by string theory, which are believed to lead to consistent, effective field theories~(EFT). However, taking into account the mentioned problem with dS vacua, and that in the string landscape it is actually easy to obtain Minkowski and Anti-de Sitter solutions, one is led to believe on the existence of the Swampland - a region wherein inconsistent semi-classical EFTs inhabit. This statement can be understood as a claim on the existence of a set of consistently-looking effective quantum field theories coupled to gravity, which are actually inconsistent with a quantum theory of gravity. And this could be an indicator that dS vacua may reside in the Swampland~\cite{Ooguri}~-~\cite{Obied}.

In this promising context, it becomes an urgent task to investigate the cosmological  implications of two of the proposed Swampland criteria, expressed as:
\begin{enumerate}
\item $SC1$: The scalar field net excursion in reduced Planck units should satisfy the bound~\cite{Ooguri}
\begin{equation}
\frac{|\Delta \phi|}{M_{P}} < \Delta \sim O(1),
\end{equation}
\item $SC2$: The gradient of the scalar field potential is bounded by~\cite{Obied}
\begin{equation}\label{eq:SC2}
M_{P}\frac{|V^{\prime}|}{V} > c \sim O(1),
\end{equation}
\end{enumerate}  
if we consider GR with the standard matter fields in the presence of a quintessence field $\phi$ to be the effective field theory. Here, both $\Delta$ and $c$ are positive constants of order one, the prime denotes derivative with respect to the scalar field $\phi$, and $M_{P} = 1/\sqrt{8\pi G}$ is the reduced Planck mass. On the other hand, it is well known that this effective field theory admits 
solutions modeling an accelerated universe, and it is reasonable to investigate and try to understand what are the conditions to be satisfied in order not to end up in the Swampland. In this regard, the one associated with $SC2$, Eq.~(\ref{eq:SC2}), is the primarily relevant and more interesting criterion to study. The two Swampland criteria  above demand that the field traverses a larger distance, in order to have the domain of validity of the effective field theory and $CS2$ to be fulfilled. 

An investigation of the implications of the string Swampland criteria based on scalar field dark energy models \cite{Lavinia} highlights the conditions to be met in order to remain outside of the Swampland. Observational implications of future surveys on quintessence models with $V(\phi) \sim e^{-\lambda \phi}$, which impose constraints on $\lambda$, are also discussed there.  The interesting question has arisen, how tightly future surveys will be able to decide whether dark energy is a cosmological constant or not. A first analysis shows that, with the data expected from Euclid, the $\lambda$ parameter should have to fall below $0.3$, leaving only room for very small deviations of quintessence from a  cosmological constant. On the other hand, the estimation constraining $\lambda < 0.1$ shows that the necessary survey volume would need to grow by a factor of $\sim 400$, as compared to that covered by the Euclid survey. Therefore, one should expect fundamental observational limitations to lowering $\lambda$ to $\lambda < 0.1$ with near-future surveys. 

Present analyses, as the one in \cite{Lavinia}, which can be considered an extension of  \cite{Agrawal}, are being performed by using a model-by-model comparision method, in order to obtain the constraints on $\lambda$ and on the $SC2$ constant $c$. In particular, the standard eight-parameter Chevallier-Polarski-Linder~(CPL) cosmology is taken as fiducial model to fit data, and then quintessence dark energy cosmology has been choosen as the comparison model. Finally, the simplest exponential potential for the quintessence field has been considered. On the other hand, also with a model-by-model comparison method, based on the belief that the universe should be multi-feature and informative, the possibly largest upper bound on the Swampland constant $c$ has been reported recently in~\cite{Wang}. In that paper, interacting quintessence is considered  as the comparison model for dark energy, constraining a 18-parameter extension of the $\Lambda$CDM cosmology, in light of current observations. The $3 \sigma$ upper bound on the Swampland constant $c$, following from this analysis, is $1.94$. Such result would permit, for instance, 11-dimensional M-theory with a double-exponential potential to be the string-theory model for dark energy. It is interesting that using Bayesian evidence as the model selection tool, the author found that this 18-parameter multi-feature cosmological model is very strongly preferred over the $\Lambda$CDM cosmology. For more details, in relation with results on the bounds on $c$ relevant to inflation, we refer the readers to~\cite{Wang}. 

Also in \cite{Wang}, the urgent necessity to clarify several important aspects concerning previous studies, which have reported different upper bounds on $c$, has been expressed. In our opinion, these differences clearly indicate that interacting dark energy models can lead to interesting deviations from the cases with no interaction. One finds in the literature examples that show how a specific form of interaction can affect the structure formation process, or how it can give rise to an effective degree of freedom to solve the cosmological coincidence problem, and how this can be incorporated to the recently announced 21-cm anomaly. Some studies point out to the fact that the result for the Hubble parameter at $z=2.34$, reported by the BOSS experiment, is also an indication of a certain interaction between dark energy and dark matter. Moreover, we have examples where this interaction, understood as an energy transfer between them, can affect the precise type and the formation of future finite-time singularities~(see, e.g., \cite{Yoo}~-~\cite{Khurshudyan4}, for more details). In light of the above mentioned facts, the results in Ref.~\cite{Wang} indicate that the interaction between dark energy and dark matter can have indeed a strong impact on the bounds on the $SC2$, Eq.~(\ref{eq:SC2}), constant $c$. However, to convert this guess into a solid conclusion a deeper investigation is required, involving different forms of linear and non-linear, sign fixed and sign changing, interactions, as the ones considered in the recent literature.  

It should be stressed again that all results discussed above have been obtained in a model-dependent way, by performing model-to-model comparison. It is  an urgent task to understand what are the upper bounds on $c$ in a dark-energy dominated universe from different observational data in a model-independent way. Will the results change substantially? Our goal in this paper is to give an aswer to this key question. For our purposes, we will use Gaussian process techniques~(GP) and $H(z)$ data. It is well known that GPs constitute a powerful tool allowing to reconstruct the behavior of a function (and its derivatives) directly from given data~\cite{Seikel} (see also \cite{hai3}). Moreover, studies carried out in the recent literature have shown that with the GP method it is possible to reconstruct the behavior of the non-gravitational interaction between dark
energy and dark matter~(among other results). It should be noted that model-independent GP techniques depend on the covariance function~(kernel), and that the hyperparameters describing it can be estimated directly from observational data (see \cite{Seikel}~-~\cite{Yang}, to mention a few). Therefore, we do not consider any specific parametrization for, e.g., the interaction term between dark energy and dark matter, but we can reconstruct it from observational data directly, using the cosmological equations. Of course, in this case reconstruction is possible if the description of dark energy is assumed. In general, the reconstruction of a function that is interesting for our study, in the scope of a certain cosmological model, will be easy to implement if we use $H(z)$ data. This is obvious, since all cosmological quantities, after some algebra, can be eventually expressed as functions of the Hubble parameter and its  derivatives, all of which can be reconstructed directly from the $H(z)$ data. Therefore, we are able, in particular,  to model the deceleration parameter at different redshifts directly from $H(z)$, by using GP, since
\begin{equation}\label{eq:dec}
q = -1 + (1+z)\frac{H^{\prime}}{H}, 
\end{equation} 
where the prime means  derivative wrt the redshift. 

In the next section, we will consider the data to be employed in this study, describing also how can we make use of GPs to reconstruct and estimate the Swampland $c$ parameter in a model-independent way. We will demonstrate that, for the study of the problem in this fashion, we do not need to make any assumption concerning the form of the scalar-field potential or dark-energy model, nor go through any model-to-model comparision as has been done till now, e.g., in~\cite{Lavinia},~\cite{Agrawal}, and \cite{Wang}. We refer the readers to several interesting works concerning  the Swampland criteria for an inflating universe~\cite{Brahma}~-~\cite{William}, and to a recently appeared discussion on the possible types of singularites for the Swampland potential $V(\phi)\sim e^{-\lambda \phi}$,  analyzed by means of the asymptotic splitting method~\cite{Oikonomou}. On the other hand, we should mention some clarifying discussions of the Swampland criteria in two works appeared recently \cite{Lavinia2} and \cite{Akrami}.  \\ 

This paper is organized as follows. In Sect.~\ref{sec:DGP} we  present the data to be used in our analysis, discussing the strategy to be followed. In Sect.~\ref{sec:Mod} we introduce our model and obtain the equations, writen in an appropiate form, which allow to see how the reconstructed behavior of the Hubble parameter and of its derivatives up to higher-order can be used for the study of the Swampland criteria in a model-independent way, by directly    using observational data. In Sect.~\ref{sec:results} we discuss the results obtained from the reconstruction for two types of kernel functions and three different values of the Hubble parameter at $z=0$, in each case. One of the values of $H_{0}$ used in our analysis has been estimated with the GP method and using high-redshift data for $H(z)$, while the other two are taken to be the values recently reported by the Planck~\cite{Ade3} and Hubble~\cite{Riess2} missions. To finish, the conclusions and a final discussion can be found in Sect.~\ref{sec:Discussion}.

\section{Data and Gaussian Processes}\label{sec:DGP}

\begin{table}[t!]
  \centering
    \begin{tabular}{ |  l   l   l  |  l   l  l  | p{2cm} |}
    \hline
$z$ & $H(z)$ & $\sigma_{H}$ & $z$ & $H(z)$ & $\sigma_{H}$ \\
      \hline
$0.070$ & $69$ & $19.6$ & $0.4783$ & $80.9$ & $9$ \\
         
$0.090$ & $69$ & $12$ & $0.480$ & $97$ & $62$ \\
    
$0.120$ & $68.6$ & $26.2$ &  $0.593$ & $104$ & $13$  \\
 
$0.170$ & $83$ & $8$ & $0.680$ & $92$ & $8$  \\
      
$0.179$ & $75$ & $4$ &  $0.781$ & $105$ & $12$ \\
       
$0.199$ & $75$ & $5$ &  $0.875$ & $125$ & $17$ \\
     
$0.200$ & $72.9$ & $29.6$ &  $0.880$ & $90$ & $40$ \\
     
$0.270$ & $77$ & $14$ &  $0.900$ & $117$ & $23$ \\
       
$0.280$ & $88.8$ & $36.6$ &  $1.037$ & $154$ & $20$ \\
      
$0.352$ & $83$ & $14$ & $1.300$ & $168$ & $17$ \\
       
$0.3802$ & $83$ & $13.5$ &  $1.363$ & $160$ & $33.6$ \\
      
$0.400$ & $95$ & $17$ & $1.4307$ & $177$ & $18$ \\

$0.4004$ & $77$ & $10.2$ & $1.530$ & $140$ & $14$ \\
     
$0.4247$ & $87.1$ & $11.1$ & $1.750$ & $202$ & $40$ \\
     
$0.44497$ & $92.8$ & $12.9$ & $1.965$ & $186.5$ & $50.4$ \\

$$ & $$ & $$ & $$ & $$ & $$\\ 

$0.24$ & $79.69$ & $2.65$ & $0.60$ & $87.9$ & $6.1$ \\
$0.35$ & $84.4$ & $7$ &  $0.73$ & $97.3$ & $7.0$ \\
$0.43$ & $86.45$ & $3.68$ &  $2.30$ & $224$ & $8$ \\
$0.44$ & $82.6$ & $7.8$ &  $2.34$ & $222$ & $7$ \\
$0.57$ & $92.4$ & $4.5$ &  $2.36$ & $226$ & $8$ \\ 
          \hline
    \end{tabular}
    \vspace{5mm}
\caption{$H(z)$ and its uncertainty $\sigma_{H}$  in  units of km s$^{-1}$ Mpc$^{-1}$. In the upper panel, $30$ samples deduced from the differential age method. In the lower pone, $10$ samples obtained from the radial BAO method. The table is according to~\cite{Zhang}~(see also references therein, for details).}
  \label{tab:Table0}
\end{table}

In order to make the discussions in Sects.~\ref{sec:Mod} and~\ref{sec:results} more transparent for the readers, we  devote the present one to some crucial aspects related to GP. In particular, we concentrate our attention to some crucial aspects related to the GP method, providing a basic knowledge of it. Additionally, the references may serve  to find more information on the topic, with some cosmological applications existing in the recent literature. To start, we recall  that the Gaussian distribution corresponds to a random variable characterized by a mean value and a covariance. Similar to Gaussian distributions,  GPs should be understood as distributions over functions, characterized by a mean function and a covariance matrix. The key ingredient of a GP is the covariance function, which for a given set of observations can infer the relation between independent and dependent variables. In other words the GP, using the covariance function, correlates the function at different points. A number of possible choices for the covariance function exist - squared exponential, polynomial, spline, etc., to mention a few. In our studies, as first option for the covariance function we chose the commonly used squared exponential function
\begin{equation}\label{eq:kernel1}
k(x,x^{\prime}) = \sigma^{2}_{f}\exp\left(-\frac{(x-x^{\prime})^{2}}{2l^{2}} \right),
\end{equation}
where $\sigma_{f}$ and $l$ are parameters known as hyperparameters. These parameters represent the length scales in the GP. The $l$ parameter corresponds to the correlation length along which the successive $f(x)$ values are correlated, while to control the variation in $f(x)$ relative to the mean of the process we need the $\sigma_{f}$ parameter. Therefore, the covariance between output variables will be written as a function of the inputs. Another interesting issue to be mentioned is that the covariance is maximum for variables whose inputs are very close. We can see from Eq.~(\ref{eq:kernel1}), that the squared exponential function is infinitely differentiable, which is a useful property in case of constructing higher-order derivatives. However, it cannot be used, for instance, to identify and study possible singularities in the future or past, based on the data used to do the reconstruction. In this regard, GPs have limited power and cannot be used to study all types of problems of modern cosmology. 

On the other hand, following the recommendations mentioned above, coming from the other studies, and aiming to reveal all possible aspects concerning the application of GP methods to study and estimate the upper bound for $SC2$, Eq.~(\ref{eq:SC2}), we use also the so-called Matern~($\nu = 9/2$) covariance function
$$k_{M}(x,x^{\prime}) = \sigma^{2}_{f} \exp \left(-\frac{3|x-x^{\prime}|}{l} \right) $$
\begin{equation}\label{eq:kernel2}
\times \left[ 1+ \frac{3 |x-x^{\prime}|}{l} + \frac{27(x-x^{\prime})}{7l^{2}} + \frac{18|x-x^{\prime}|^{3}}{7l^{3}} + \frac{27 (x-x^{\prime})^{4}}{35 l^{4}}\right], 
\end{equation}
 
In the following we will use the publicly available package GaPP~(Gaussian Processes in Python) developed by Seikel et al. It allows to choose different covariance functions, including the Matern covariance function given by Eq.~(\ref{eq:kernel2}), while the squared exponential function, Eq.~(\ref{eq:kernel1}), is used in the code as a default option. The code is also very useful to combine different observational datasets, provided the proper relation between them is known. Below, we present the used dataset and some clarification about the accepted strategy for our study, concerning the Hubble parameter value at $z=0$. In particular, we use $30$-point samples of $H(z)$ deduced from the differential age method. Then, we add $10$-point samples obtained from the radial BAO method (see Table~\ref{tab:Table0}). In the first case, as we can see from Table~\ref{tab:Table0}, we have relatively good data up to $z=2$. On the other hand, the added data-points from the radial BAO method allow us to extend the data range up to $z = 2.4$ improving also low-redshift data. However, we can see that the presented data-points in Table~\ref{tab:Table0} do not include the value of the Hubble parameter at $z=0$, i.e. the value of  $H_{0}$. This value will play an important role in our study,  what will be seen in Sect.~\ref{sec:Mod} at the theoretical level, and in Sect.~\ref{sec:results} during the discussion of the results. It should be noted that in our study we consider three different values for $H_{0}$. In particular, for two cases, we considered the value of $H_{0}$ reported by the Planck and the Hubble missions, respectively, while, separately,  we will allow the GP itself to estimate $H_{0}$ using the values of the $H(z)$ data-points from higher redshifts, presented in Table~\ref{tab:Table0}. 

\begin{figure}[h!]
 \begin{center}$
 \begin{array}{cccc}
\includegraphics[width=150 mm]{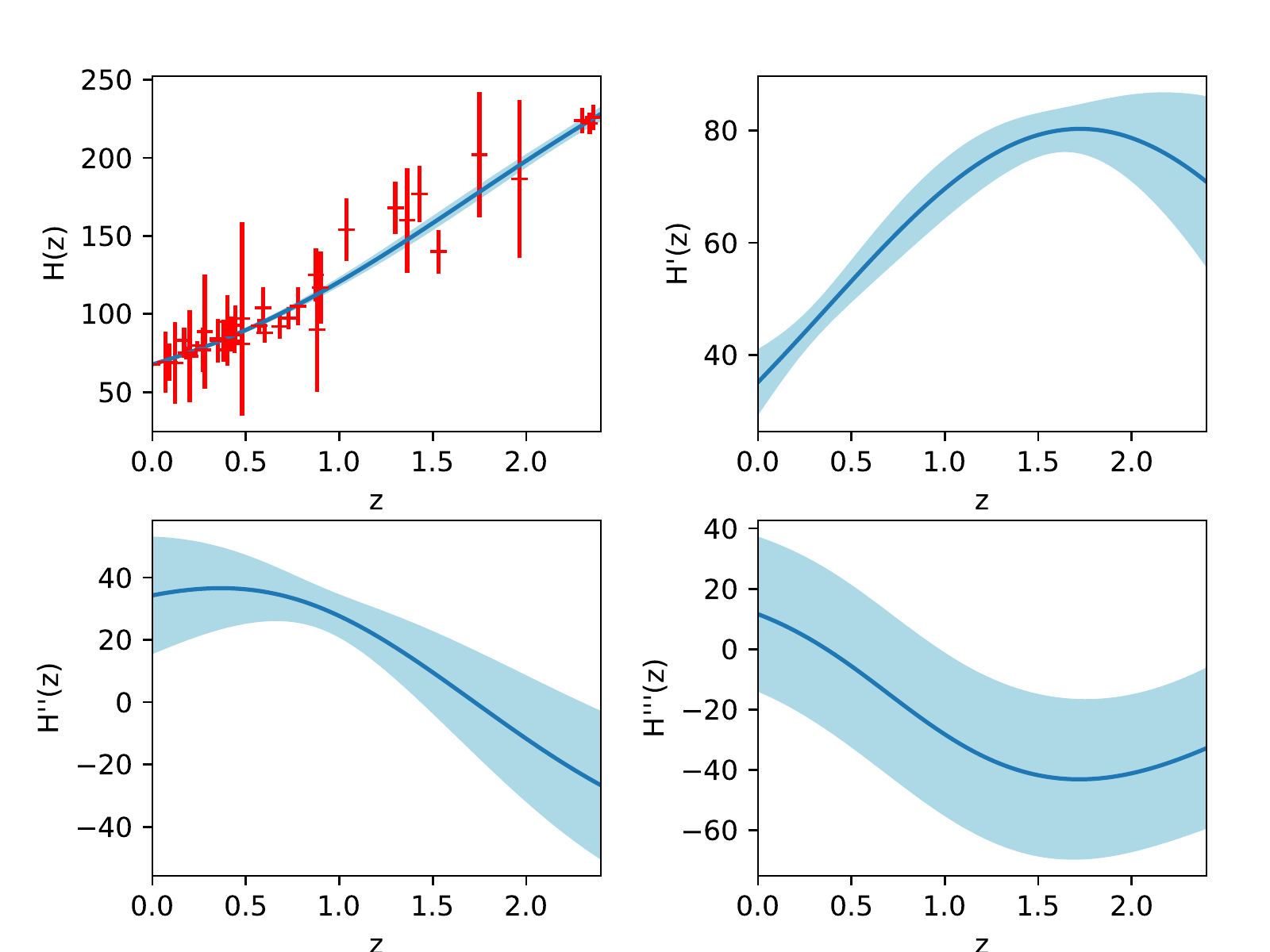} &&
 \end{array}$
 \end{center}
\caption{GP reconstruction of $H(z)$, $H^{\prime}(z)$, $H^{\prime \prime}(z)$, and $H^{\prime \prime \prime}(z)$, for the $40$-point sample deduced from the differential age method, with the additional 10-point sample obtained from the radial BAO method, when $H_{0} = 67.66 \pm 0.42$ reported by the Planck mission. The $^{\prime}$ means derivative with respect to the redshift variable $z$.}
 \label{fig:Fig0_1}
\end{figure}

\begin{figure}[h!]
 \begin{center}$
 \begin{array}{cccc}
\includegraphics[width=150 mm]{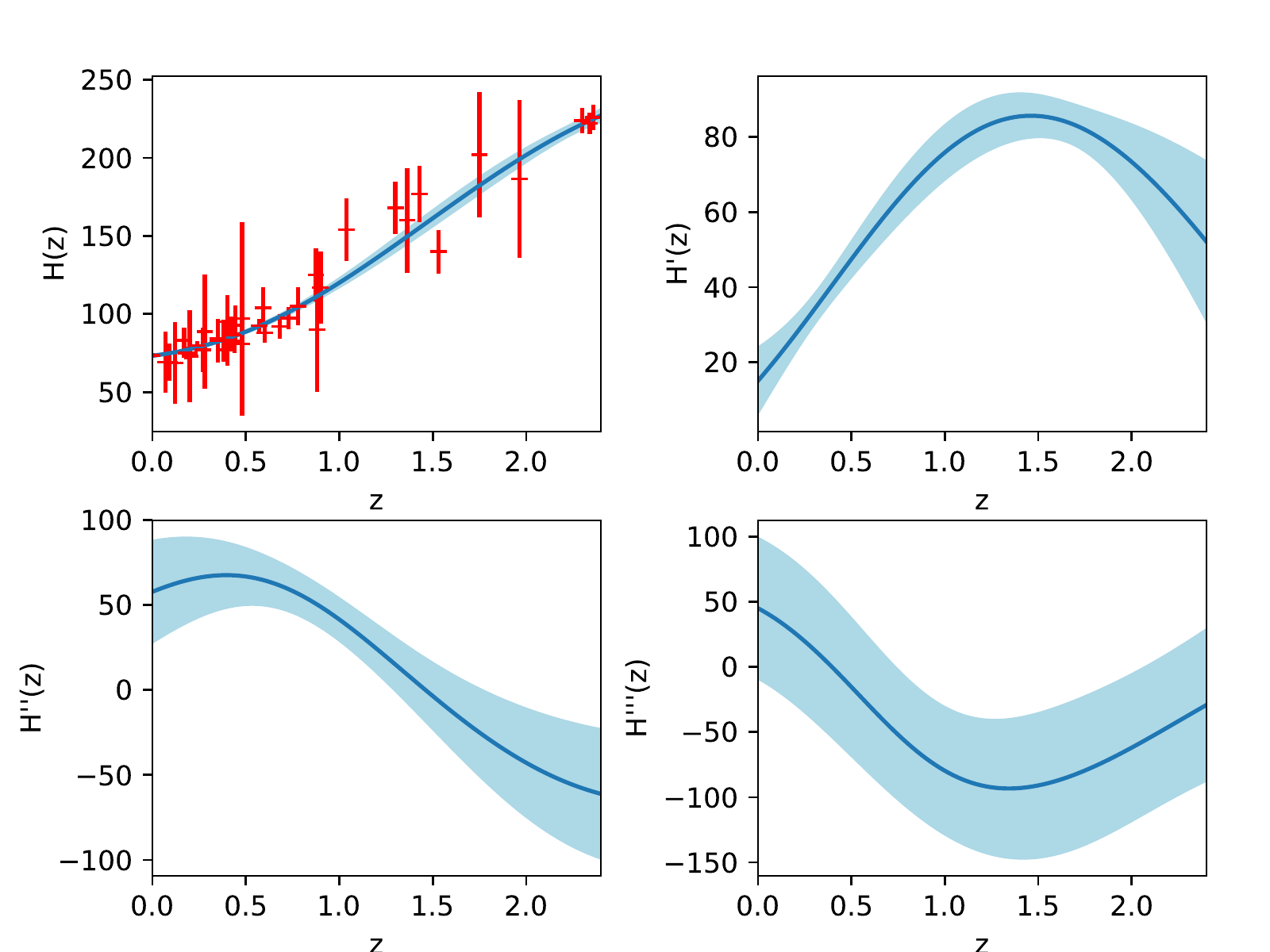} &&
 \end{array}$
 \end{center}
\caption{GP reconstruction of $H(z)$, $H^{\prime}(z)$, $H^{\prime \prime}(z)$ and $H^{\prime \prime \prime}(z)$ for the $40$-point sample deduced from the differential age method, with the additional 10-point sample obtained from the radial BAO method, when $H_{0} = 73.52 \pm 1.62$ reported by the Hubble mission. The $^{\prime}$ means derivative with respect to the redshift $z$.}
 \label{fig:Fig0_2}
\end{figure}

A situation similar to the last case has been already used, recently, in a study dedicated to a new dark energy parametrization, given by $\omega = \omega_{0} + \omega_{1} q$, where $\omega_{0}$ and $\omega_{1}$ are the parameters of the model, to be determined, while $q$ is the deceleration parameter Eq.~(\ref{eq:dec}). To save space, let us refer the readers to~\cite{Emilio}, where the value of $H_{0}$ has been estimated for two cases using GP directly. In particular, the authors found that the GP can estimate it, yielding $H_{0} = 71.286 \pm 3.743$ and $67.434 \pm 4.748$~(at $1\sigma$ reconstruction level) for $40$ and $30$-point samples of $H(z)$ data, respectively. On the other hand, in the same work the authors presented the reconstructed behavior of the Hubble parameter and its higher-order derivatives for the squared exponent kernel given by Eq.~(\ref{eq:kernel1}) \cite{Emilio}. Here we will present only the results of the reconstruction for $H_{0} = 67.66 \pm 0.42$ and $H_{0} = 73.52 \pm 1.62$ reported from the Planck and Hubble missions, respectively, for the Matern~($\nu = 9/2$) kernel, Eq.~(\ref{eq:kernel2}); see Figs.~(\ref{fig:Fig0_1}) and~(\ref{fig:Fig0_2}), which show the results of the reconstruction.

To end this section, observe that a visual comparison of the behavior of the reconstructed $H$ and $H^{\prime}(z)$ indicates some differences between the considered cases, which could leave some effect on the estimations under study. In the next section we will see that, for the model independent way to estimate the upper bound of $SC2$, we need just the reconstructed behavior of the Hubble parameter and of its first-order derivative, when $H_{0}$ and $\Omega_{dm}$ at $z=0$ are known. More on this issue in Sect.~\ref{sec:results}. In the next, we present a detailed  demonstration on how the results of this one can be applied on the the problem under study.

\section{Model}\label{sec:Mod}

Here we shall consider GR with the standard matter field in the presence of a quintessence field $\phi$ to be the EFT described by the following action
\begin{equation}
S = \int d^{4} x \sqrt{-g} \left( \frac{M^{2}_{P}}{2} R - \frac{1}{2} \partial_{\mu} \phi \partial^{\mu} \phi -V(\phi) \right ) + S_{m},
\end{equation}
where $S_{m}$ corresponds to standard matter, $M_{P} = 1/\sqrt{8\pi G}$ is the reduced Planck mass, $R$ is the Ricci scalar, $\phi$ the field, and $V(\phi)$ is the field potential. It is well known that,  when we consider the FRWL universe, the dynamics of the scalar field's dark energy and dark matter will be described by the  equations 
\begin{equation}\label{eq:drhoPi}
\dot{\rho}_{\phi} + 3 H (\rho_{\phi} + P_{\phi}) = 0,
\end{equation}
\begin{equation}\label{eq:drhoDm}
\dot{\rho}_{dm} + 3 H \rho_{dm} = 0.
\end{equation}
In other words, Eqs.~(\ref{eq:drhoPi}) and~(\ref{eq:drhoDm}) are the energy conservation  laws for the components describing the background dynamics. Moreover, the form of these equations demonstrates the absence of a coupling between the scalar field's dark energy and dark matter, accounted for in the recent literature as an energy flow between them. The presence of this coupling is interpreted as an interaction between dark energy and dark matter. Furthermore, we know that $\rho_{\phi}$, $\rho_{dm}$ and $P = P_{\phi}$ are related to each other through the Friedmann equations, as follows
\begin{equation}\label{eq:F1}
H^{2} = \frac{1}{3} (\rho_{\phi} + \rho_{dm}), 
\end{equation}
and 
\begin{equation}\label{eq:F2}
\dot{H} + H = -\frac{1}{6} (\rho_{\phi} + \rho_{dm} + 3 P_{\phi}).
\end{equation}

On the other hand, assuming that the scalar field is spatially homogeneous for its energy density and pressure we have
\begin{equation}\label{eq:rhoPhi}
\rho_{\phi} = \frac{1}{2} \dot{\phi} + V(\phi),
\end{equation}
and
\begin{equation}\label{eq:pPhi}
P_{\phi} = \frac{1}{2} \dot{\phi} - V(\phi),
\end{equation}
where the dot is the derivative w.r.t to the cosmic time, while $V(\phi)$ is the potential of the scalar field. In all equations  above $H = \dot{a}/a$ is the Hubble parameter.

Now, some basic aspects concerning the background dynamics in the presence of scalar-field dark energy and standard matter: let us see how one can involve GP in the study of such models in a model-independent way. In particular, how the $H(z)$ data under  consideration can be used, eliminating the need to have the form of the scalar field potential given in advance. In particular, it is easy to see, from Eqs.~(\ref{eq:rhoPhi}) and~(\ref{eq:pPhi}), that
\begin{equation}\label{eq:dphi2}
\dot{\phi}^{2} = \rho_{\phi} + P_{\phi},
\end{equation} 
while
\begin{equation}\label{eq:Vphi}
V(\phi) = \frac{\rho_{\phi} - P_{\phi}}{2}.
\end{equation}    
Now, as from Eq.~(\ref{eq:drhoDm}) we have $\rho_{dm} = 3 H_{0}^{2} \Omega_{0} (1+z)^{3}$, then from Eq.~(\ref{eq:F1}) we can determine the energy density of the scalar field, which in this case reads as follows
\begin{equation}
\rho_{\phi} = 3 H^{2} - 3 H_{0}^{2} \Omega_{0} (1+z)^{3},
\end{equation}
where $H_{0}$ is the Hubble parameter value at $z=0$~(z is the redshift). It is then clear that, in order to perform the analysis of the model and estimate the upper bound of $SC2$, we need to determine the functional dependence of $P_{\phi}$ on $H$. Which is an easy task and can be done using Eq.~(\ref{eq:F2}). After some algebra, we see that  $P_{\phi} = 2(1+z)H H^{\prime} - 3 H^{2}$, where the prime denotes  derivative wrt the redshift. Of course, we can see immediately that $\rho^{\prime}_{\phi} = 6 H H^{\prime} - 9  H_{0}^{2} \Omega_{0} (1+z)^{2}$ and $P^{\prime}_{\phi} = 2 (1+z) (H^{\prime 2} + H H^{\prime \prime}) - 4 H H^{\prime}$. Coming back to the form of $SC2$ to be reconstructed, we need only take into account that $d V(\phi)/d\phi = (dV/dz)/(d\phi/dz)$, where $d\phi/dz$ should be calculated from Eq.~(\ref{eq:dphi2}), and that $\dot{\phi} = -(1+z)H \phi^{\prime}$. 

After all this,  it has become clear that we are in the position to reconstruct $SC2$ and  estimate its upper bound in a model-independent way using directly observational data. The results of the study, for the strategies discussed in Sect.~\ref{sec:DGP} are presented in the next section.

\section{Results}\label{sec:results}

In section~\ref{sec:DGP} we already mentioned that the analysis will be done for two kernel functions for three different values of the Hubble parameter at $z=0$. Let us start the discusion of the results obtained from the first case, corresponding to the squared-exponential kernel function, Eq.~(\ref{eq:kernel1}), with the value of the Hubble parameter coming from the GP from higer-redshift data  Table~\ref{tab:Table0}. In this case we have seen that, according to the mean value of the reconstruction, $H_{0} \approx 71.28$, while according to  the $1\sigma$ reconstruction, the $1\sigma$ error is $\approx 3.74$~\cite{Emilio}. On the other hand, the reconstruction of $\Omega_{\phi} = \rho_{\phi}/3H^{2}$ shows that the model should be rejected above $z \approx 1.9$ since the mean of the reconstruction predicts a negative $\Omega_{\phi}$. Moreover, we see also that, according to the reconstructed behavior of the mean, the dark energy dominated universe will be observed from $z\approx 0.27$, while according to the $2\sigma$ reconstruction band the dark energy dominated epoch will start from $z \approx 0.5$. On the other hand, we also have been able to estimate $\Omega_{\phi}$ at $z=0$ giving $\Omega_{\phi} \approx 0.7^{+ 0.05 + 0.08}_{-0.05 - 0.11}$ according to the mean and $1\sigma$ and $2\sigma$ of the reconstruction bands, respectively.

\begin{figure}[h!]
 \begin{center}$
 \begin{array}{cccc}
\includegraphics[width=90 mm]{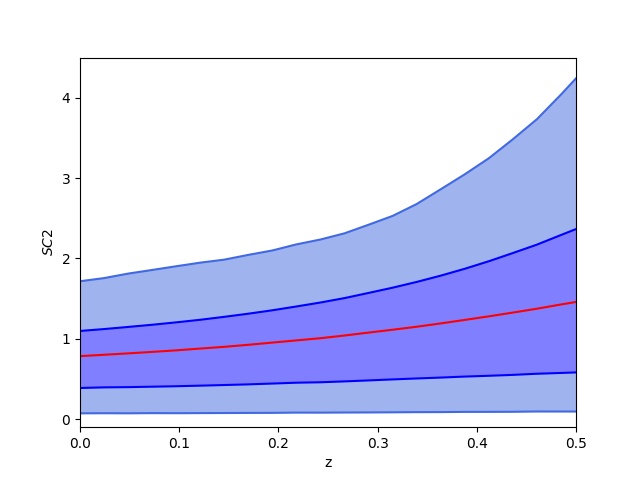} &&
\includegraphics[width=90 mm]{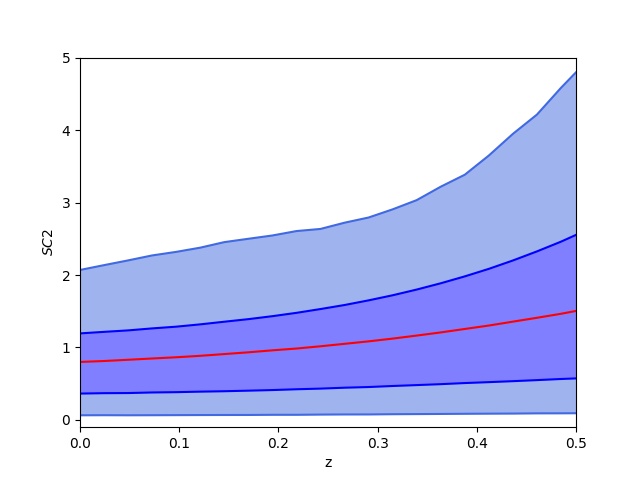} \\
 \end{array}$
 \end{center}
\caption{Reconstruction of the $|V^{\prime}|/V$, Eq.~(\ref{eq:SC2}), from the
$H(z)$ data depicted in Table~\ref{tab:Table0}. The left panel
represents GP reconstruction for the squared exponent kernel given by Eq.~(\ref{eq:kernel1}), while the right plot has been obtained considering the  Matern~$(\nu = 9/2)$  kernel given by Eq.~(\ref{eq:kernel2}). The solid line is the mean of the reconstruction and the shaded blue regions are the $68\%$ and $95\%$ C.L. of the reconstruction, respectively. $H_{0} = 71.286 \pm 3.743$ has been estimated by GP using the data in Table~\ref{tab:Table0}.}
 \label{fig:Fig1}
\end{figure}

It should be mentioned as well that the results obtained for $2\sigma$ in future could be questionable, since the estimation of $\Omega_{\phi}$ from the reconstruction induces a tension, giving result that are not consistent with the results from the other missions. However, for this case we also estimated the upper bound on the constant $c$ of $SC2$, Eq.~(\ref{eq:SC2}). This could already be an indicator that we cannot  trust too much the results from the reconstructed $SC2$, unless new data are available to allow such possibility, for the low redshift universe. The two plots of Fig.~(\ref{fig:Fig1}) correspond to  the reconstruction of $SC2$ allowing to estimate its upper bound, obtained by involving model-independent processes, as discussed above. 

The left plot of Fig.~(\ref{fig:Fig1}) corresponds to  the reconstruction of $SC2$ for the squared exponent case, Eq.~(\ref{eq:kernel1}), while the reconstruction corresponding to the Matern~$(\nu = 9/2)$ kernel given by Eq.~(\ref{eq:kernel2}) can be found on the rhs plot. From these plots we see, that the GP and $H(z)$ data presented in Table~\ref{tab:Table0} yield a quite good reconstruction of $SC2$, allowing to obtain the upper bounds on this parameter according to the mean, $1\sigma$, and $2\sigma$ reconstructed bounds, respectively. The results of a further analysis show that:  

\begin{enumerate}

\item According to the mean of the reconstruction, in the case of the squared exponent kernel, Eq.~(\ref{eq:kernel1}), the dark energy dominated universe should start from $z\approx 0.27$, while we will observe a dark energy dominated universe from $z\approx 0.37$ according to the upper bound of the $1\sigma$ reconstruction. On the other hand, from $z\approx 0.5$ we can observe a dark energy dominated universe if we take into account the upper band from the $3\sigma$ reconstruction. The same picture has been observed after the reconstruction where we used the Matern~$(\nu = 9/2)$ kernel given by Eq.~(\ref{eq:kernel2}). Moreover, at $z=0$, we will have $\omega_{de} \approx -1.15$, $\omega_{de} \approx -0.96$ and $\omega_{de} \approx -0.76$, from the mean and the upper bounds for the $1\sigma$ and $2\sigma$ reconstructions, respectively. However, when we start the estimation of $\omega_{de}$ at $z=0$, we observe that $\omega_{de} \approx - 1.13$, $\omega_{de} \approx -0.932$, and $\omega_{de} \approx -0.705$, from the mean and the upper bounds for the $1\sigma$ and $2\sigma$ reconstructions, respectively.

\item On the other hand, according to the mean of the reconstruction, for the upper bound on the $SC2$ constant $c$ for $z$ $z\in [0,0.27]$, we will have $\approx 0.785$. Moreover, according to the $1\sigma$ reconstruction, the upper bound on $c$ for $z\in [0,0.37]$ will be $\approx 1.786$, while the $\approx 4.363$ upper bound for $c$ will be observed from the $3\sigma$ reconstruction bands. This estimation has been obtained with the squared exponent kernel, Eq.~(\ref{eq:kernel1})~(see the left plot of Fig.~(\ref{fig:Fig1})). When we use the Matern~$(\nu = 9/2)$ kernel, Eq.~(\ref{eq:kernel2}), we observe that the upper bounds on $c$ will be $\approx 1.051$, $\approx 1.886$, and $\approx 4.926$, respectively, as it has been discussed for the previous case~(see the rhs plot of Fig.~(\ref{fig:Fig1})).

\end{enumerate}

\begin{figure}[h!]
 \begin{center}$
 \begin{array}{cccc}
\includegraphics[width=90 mm]{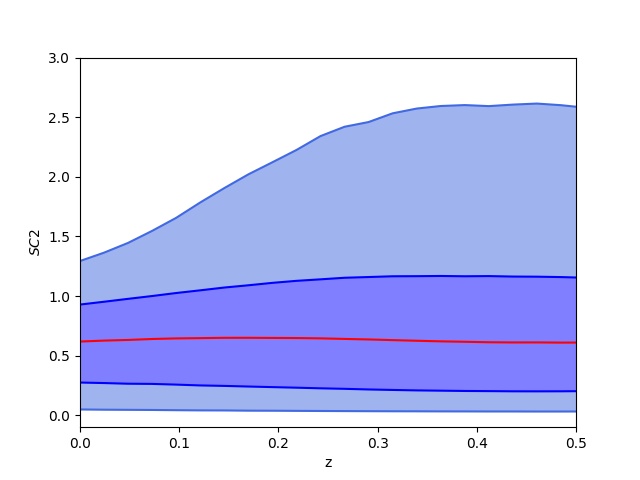} &&
\includegraphics[width=90 mm]{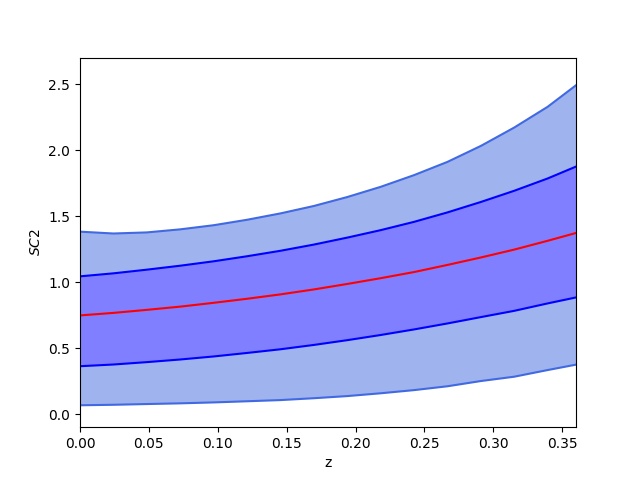} \\
 \end{array}$
 \end{center}
\caption{Reconstruction of the $|V^{\prime}|/V$, Eq.~(\ref{eq:SC2}), from the
$H(z)$ data depicted in Table~\ref{tab:Table0}. The left panel
corresponds to the GP reconstruction for the squared exponent kernel given by Eq.~(\ref{eq:kernel1}), while the rhs plot has been obtained considereng the kernel to be Matern~$(\nu = 9/2)$, given by Eq.~(\ref{eq:kernel2}). The solid line is the mean of the reconstruction and the shaded blue regions are the $68\%$ and $95\%$ C.L. of the reconstruction, respectively. $H_{0} = 73.52 \pm 1.62$ has been used according to the Hubble mission result~\cite{Riess2}.}
 \label{fig:Fig2}
\end{figure}

On the other hand, the results obtained from the case when we consider $H_{0} = 73.52 \pm 1.62$ reported in~\cite{Riess2} are summarized below and are presented in Fig.~(\ref{fig:Fig2}). In particular during the study we observed, that

\begin{enumerate}

\item  According to the mean of the reconstruction in case of squared exponent kernel, Eq.~(\ref{eq:kernel1}), the dark energy dominated universe should start from $z\approx 0.34$, while we will observe dark energy dominated universe from $z\approx 0.42$ according to the upper bound of $1\sigma$ reconstruction. On the other hand, from $z\approx 0.5$ we can observe dark energy dominated universe if we take into account the upper band from $3\sigma$ reconstruction. However, if the reconstruction is done considering Matern~$(\nu = 9/2)$ kernel given by Eq.~(\ref{eq:kernel2}), then the dark energy dominated universe will be observed from $z\approx 0.27$, $z\approx 0.31$ and $z \approx 0.37$ for the mean and the upper bands of $1\sigma$ and $2\sigma$ reconstruction, respectively. Moreover, at $z=0$, we will have $\omega_{de} \approx -0.96$, $\omega_{de} \approx -0.88$ and $\omega_{de} \approx -0.81$ from the mean and the upper bounds from $1\sigma$ and $2\sigma$ reconstructions respectively when the kernel is given by Eq.~(\ref{eq:kernel1}). On the other hand, when we considered the kernel given by Eq.~(\ref{eq:kernel2}), for the estimation of $\omega_{de}$ at $z=0$, we observed that $\omega_{de} \approx - 1.23$, $\omega_{de} \approx -1.12$ and $\omega_{de} \approx -0.99$ from the mean and the upper bounds from $1\sigma$ and $2\sigma$ reconstructions, respectively.

\item On the other hand, according to the mean of the reconstruction, the upper bound on the $SC2$ constant $c$ for $z$ $z\in [0,0.34]$ we will have $\approx 0.649$. On the other hand, according to the $1\sigma$ reconstruction the upper bound on $c$ for $z\in [0,0.42]$ will be $\approx 1.167$, while $\approx 2.61$ upper bound for $c$ will be observed from the $3\sigma$ reconstruction bands on $z\in [0,0.5]$. The presented estimation has been obtained with squared exponent kernel, Eq.~(\ref{eq:kernel1})~(see the left plot of Fig.~(\ref{fig:Fig2})). On the other hand, when we used Matern~$(\nu = 9/2)$ kernel, Eq.~(\ref{eq:kernel2}), we observed that the upper bounds on $c$ will be $\approx 1.129$, $\approx 1.691$ and $\approx 2.52$, respectively~(see the right plot of Fig.~(\ref{fig:Fig2})).

\end{enumerate}

\begin{figure}[h!]
 \begin{center}$
 \begin{array}{cccc}
\includegraphics[width=90 mm]{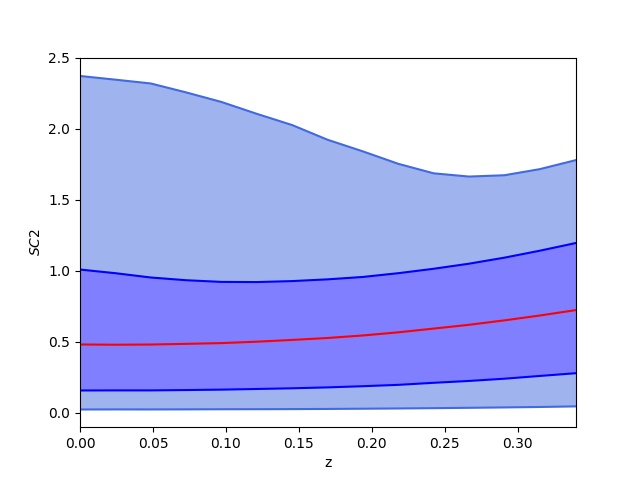} &&
\includegraphics[width=90 mm]{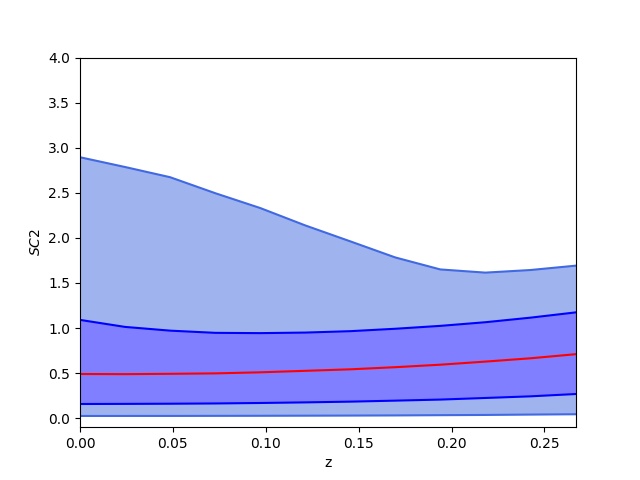} \\
 \end{array}$
 \end{center}
\caption{Reconstruction of  $|V^{\prime}|/V$, Eq.~(\ref{eq:SC2}), from the
$H(z)$ data  in Table~\ref{tab:Table0}. The left panel
corresponds to GP reconstruction for the squared exponent kernel given by Eq.~(\ref{eq:kernel1}), while the right plot comes from considering the kernel to be Matern~$(\nu = 9/2)$, as given by Eq.~(\ref{eq:kernel2}). The solid line depicts the mean of the reconstruction and the shaded blue regions are the $68\%$ and $95\%$ C.L. of the reconstruction, respectively. The value $H_{0} = 67.66 \pm 0.42$ has been used, in accordance with the latest Planck  results~\cite{Ade3}.}
 \label{fig:Fig3}
\end{figure}

Finaly, we would like to sumarize the results obtained from the study when we assume that the value of the Hubble parameter comes from the Planck mission result, i.e. $H_{0} = 67.66 \pm 0.42$~\cite{Ade3}. The results  for both kernel functions can be summarized as follows:

\begin{enumerate}

\item According to the mean of the reconstruction in the case of the squared exponent kernel, Eq.~(\ref{eq:kernel1}), the dark energy dominated universe should start from $z\approx 0.15$, while we will observe a dark energy dominated universe from $z\approx 0.25$, according to the upper bound of the $1\sigma$ reconstruction. On the other hand, from $z\approx 0.36$, we can observe a dark energy dominated universe if we take into account the upper band from the $3\sigma$ reconstruction. However, if the reconstruction is done considering the Matern~$(\nu = 9/2)$ kernel given by Eq.~(\ref{eq:kernel2}), then the dark energy dominated universe will be observed from $z\approx 0.09$, $z\approx 0.19$ and $z \approx 0.27$ for the mean and the upper bands of the $1\sigma$ and $2\sigma$ reconstruction, respectively. Moreover, at $z=0$, we will have $\omega_{de} \approx -1.097$, $\omega_{de} \approx -0.977$ and $\omega_{de} \approx -0.872$ from the mean and the upper bounds from the $1\sigma$ and $2\sigma$ reconstructions, respectively, when the kernel is given by Eq.~(\ref{eq:kernel1}). On the other hand, when we consider the kernel given by Eq.~(\ref{eq:kernel2}), for the estimation of $\omega_{de}$ at $z=0$, we observe that $\omega_{de} \approx - 1.15$, $\omega_{de} \approx -1.011$ and $\omega_{de} \approx -0.894$, from the mean and the upper bounds from the $1\sigma$ and $2\sigma$ reconstructions, respectively.

\item On the other hand, according to the mean of the reconstruction, for the upper bound on the $SC2$ constant $c$ for $z$ $z\in [0,0.15]$, we will have $\approx 0.52$. Moreover, according to the $1\sigma$ reconstruction, the upper bound on $c$ for $z\in [0,0.25]$ will be $\approx 1.012$, while the $\approx 2.37$ upper bound for $c$ will be observed from the $3\sigma$ reconstruction bands on $z\in [0,0.36]$. The  estimation has been obtained with the squared exponent kernel, Eq.~(\ref{eq:kernel1})~(see the left plot of Fig.~(\ref{fig:Fig3})). Finally, when we use the Matern~$(\nu = 9/2)$ kernel, Eq.~(\ref{eq:kernel2}), we observe that the upper bounds on $c$ are $\approx 0.51$, $\approx 1.02$ and $\approx 2.895$, respectively~(see the rhs plot of Fig.~(\ref{fig:Fig3})).

\end{enumerate}

In summary, analysis of this paper allows to estimate the upper bound on the constant of $SC2$ using GP and $40$-point $H(z)$ data. This is a fully model-independent estimation, since we do not include any assumption on the form of the Swampland potential, and we do not preclude any model to compare, in order to do the estimation. According to the discussion presented above, we found that, when $H_{0} = 67.66 \pm 0.42$, according to the Planck results, then the upper bounds on the $SC2$ constant are $\approx 2.37$ and $\approx 2.895$, for the squared exponent and Matern~$(\nu = 9/2)$ kernel, respectively. On the other hand, when $H_{0} = 73.52 \pm 1.62$, then the model-independent estimation of the upper bounds are found to be $\approx 1.167$, according to the $1\sigma$ reconstruction. On the other hand, when we use the Matern~$(\nu = 9/2)$ kernel, Eq.~(\ref{eq:kernel2}), we observe  the upper bound to be $\approx 1.691$. Moreover, when $H_{0} = 71.286 \pm 3.743$ is considered, the upper bound on $SC2$ turns to be $\approx 0.785$. Further, according to the $1\sigma$ reconstruction the upper bound on $c$ reads $\approx 1.786$~(for the squared kernel function). However, when we use the Matern~$(\nu = 9/2)$ kernel, Eq.~(\ref{eq:kernel2}), then we observe that the upper bounds on $c$ read $\approx 1.051$ and $\approx 1.886$, respectively.       

The above  estimations of the upper bound of the constant $c$ of $SC2$, Eq.~(\ref{eq:SC2}), prove the possibility to have different upper bounds depending on the kernel function considered, and on the present-day value of the Hubble parameter. This is a very interesting result, which non-trivially complements the upper bounds reported by other(model-dependent) studies in the recent literature. In particular, our results are perfectly consistent with the results reported in~\cite{Wang}. But, on the other hand, we see that, in a model-independent analysis as the one performed here, we can even reach upper bounds behind these, what had been advanced as a feasible possibility in previous studies. Our results here indicate, in our opinion, that each detail concerning the way the fit is performed, the background dynamics in model-to-model comparison methods adopted in previous studies, including the priors and used datasets, can significantly affect the final values obtained for the upper bounds on $c$. Moreover, with the GP estimation of the upper bound on $c$  adopted here, in order to reject or recover the status of EFT theories one still requires additional analysis considering other datasets and kernels with different priors on the hyperparameters. The study here performed should be viewed as just a first (albeit already fruitful) attempt to use GP methods to study Swampland criteria for the dark energy dominated universe.

\section{\large{Discussion and conclusions}}\label{sec:Discussion}

In this paper, we have used GP techniques in order to investigate the implications of the string Swampland criteria for a scalar-field dark-energy dominated universe, without assuming any prior specific form for the field potential. In other words, we have considered GR, with a standard matter field in the presence of a quintessence field, $\phi$, without fixing the field potential, to be the effective field theory. Our study consists in a fully model-independent analysis: we invoke GP to reconstruct the form of the potential from $H(z)$ data, and estimate at the same time the upper bound on the constant $c$ of $SC2$. The $40$-points $H(z)$ data used in the process consists of $30$-point samples coming from the differential age method, and additional $10$-point samples obtained from the radial BAO method. 

The upper bounds on the second Swampland Criteria~($|V^{\prime}|/V$) have been estimated both for the squared exponent and Matern~($\nu = 9/2$) kernels, for three different values of $H_{0}$ in each case. Specifically, in one case we have made use of the GP working with the $H(z)$ data sample to estimate the value of $H_{0}$, while in the other two situations, we have considered  values of $H_{0}$ compatible with those reported by the Planck and Hubble missions, i.e., $H_{0} = 67.66 \pm 0.42$, and $H_{0} = 73.52 \pm 1.62$, respectively. After some algebra we have found that, in the absence of a coupling between the scalar field's dark energy and dark matter, we can express $SC2$, Eq.~(\ref{eq:SC2}), in terms of $H$ and $H^{\prime}$, which are reconstructed by means of the GP. After establishing a proper mathematical background, we were able to estimate the upper bounds on $SC2$, for each case, in a completely model-independent way. It is obvious that this approach is quite different from previous model-to-model comparisons; however, the estimations do heavily depend on the quality of the data used for the reconstruction. Eventually,  estimations also rely on the kernel used, which not only controls the mean value of the reconstruction, but also the error bars of the same, and the reconstruction of the derivatives of the $H$ parameter, as well. We should stress once more that previous estimations of the upper bounds on the constant of $SC2$ were based on methods involving model-to-model comparison, while here we have merely used a kernel and performed the estimation from the best observational data available.

In parallel to the reconstruction of $SC2$, and in order to estimate the upper bounds on its value, we have also reconstructed $\Omega_{de}$, $\omega_{de}$ and could estimate at which redshifts the dark energy dominated universe can be observed, in order to address the $SC2$ estimation. In particular, we have concluded that, when we involve the GP alone to estimate the Hubble parameter value at $z=0$ (found to be $H_{0} = 71.286 \pm 3.743$), then,  according to the mean value of the reconstruction, in the case of the squared exponent kernel, the dark energy dominated universe starts at $z\approx 0.27$. On the other hand, the dark energy dominated universe will start from $z\approx 0.37$, when the upper bound of the $1\sigma$ reconstruction is used. Finally, in the case  we take into account the upper band corresponding to the $3\sigma$ reconstruction, we get a dark-energy dominated universe from $z\approx 0.5$ onward. 

A similar picture has been obtained, as a result of the reconstruction procedure, when we have used the Matern~$(\nu = 9/2)$ kernel, given by Eq.~(\ref{eq:kernel2}). The reconstruction of $\Omega_{\phi} = \rho_{\phi}/3H^{2}$ shows that the model should be rejected above $z \approx 1.9$, since the mean of the reconstruction predicts a negative $\Omega_{\phi}$. Eventually, we were also able to estimate $\Omega_{\phi}$ at $z=0$, yielding $\Omega_{\phi} \approx 0.7^{+ 0.05 + 0.08}_{-0.05 - 0.11}$, according to the mean and $1\sigma$ and $2\sigma$ results for the reconstruction bands, respectively. The above analysis yields the following results for the squared kernel function: according to the mean of the reconstruction, the upper bound on the $SC2$ is $\approx 0.785$, for the $1\sigma$ reconstruction, the upper bound on $c$ is $\approx 1.786$, while for the $3\sigma$ reconstruction, we get $\approx 4.363$ as upper bound for $c$. Alternatively, when we used the Matern~$(\nu = 9/2)$ kernel, Eq.~(\ref{eq:kernel2}), we obtained that the upper bounds on $c$ turn out to be $\approx 1.051$, $\approx 1.886$, and $\approx 4.926$, respectively.  Surprisingly, in both cases  higher upper bounds have been obtained for the $3\sigma$ reconstruction bounds, in which case $\Omega_{\phi} \approx 0.782$, for both kernel functions. Therefore, the most reliable outcome,  under the form of an upper bounds for the $c$ constant of $SC2$, appears to be $\approx 1.786$ and $\approx 1.886$, obtained from the $2\sigma$ reconstruction bounds for the two kernels, respectively. 

In addition, considering $H_{0} = 73.52 \pm 1.62$ and estimating $\Omega_{\phi}$ at $z=0$, we are led to $\Omega_{\phi} \approx 0.74$, for the upper bound of the $3\sigma$ reconstruction. Again in this case, similarly to the first one discussed above, we better trust the results obtained from the mean and $1\sigma$ reconstruction bands. In this regard, according to the mean value of the reconstruction, the upper bound lies at $\approx 0.649$, while it rises to $\approx 1.167$, according to the $1\sigma$ reconstruction. Alternatively, when using the Matern~$(\nu = 9/2)$ kernel, Eq.~(\ref{eq:kernel2}), we observed that the upper bounds on $c$ turn to be $\approx 1.129$ and $\approx 1.691$, for the mean and $2\sigma$ reconstruction bands, respectively.

Finally, the analysis of the case when $H_{0} = 67.66 \pm 0.42$ has been performed. It reveals that, as more reliable upper bounds on the constant of $SC2$, those  obtained from $1\sigma$ and $2\sigma$ reconstructions should be taken, because only for them can one obtain results for $\Omega_{\phi}$ and $\omega_{\phi}$ that are consistent with the results reported by other studies. In this case, for the upper bound for $SC2$ we get $\approx 1.012$ and $\approx 2.37$, when the squared exponent kernel is considered. If we start from the Matern~$(\nu = 9/2)$ kernel, the consistent background dynamics can be observed when using the $1\sigma$ and $2\sigma$ reconstruction upper bands, namely $\approx 1.02$ and $\approx 2.895$, respectively.

As argued above, the dS solution seems to be in the Swampland, what would  rule out $\Lambda$CDM in the future of the universe and maybe start to generate some tensions at present. More specifically, it has been argued  in~\cite{colg3} that $H(z)$ ought to have a turning point at some low value of $z$; but  the results of our analysis do not seem to show such implication. It appears as if, at the level of our present research, the Swampland conjecture could be disproven. However, it would not be reasonable to adventure such  result with only one case considered; a more rigorous analysis using different datasets when involving the GP must be undertaken in order to be able to reach such sharp and important conclusion. We expect to return to this relevant point soon, by increasing the accuracy of our analysis.
       
To summarize, our model-independet estimations for $SC2$ are in good agreement with the results reported in~\cite{Wang} for the dark energy dominated universe. However, we have noticed the possibility to get higher upper bounds on the $SC2$ constant, never reported before. This, probably, could be achieved directly, by using appropriate forms for the interaction term between  the scalar field and dark matter. This may be, of course, a hard task to perform, since there are various possible forms for the interaction term and checking any of them is a very time-consuming process. In this regard, using GP techniques can again be very useful. The study reported here indicates that every detail, concerning the way the fit was performed, and the background dynamics in the model-to-model comparison method adopted in previous studies, including the priors and datasets used, can significantly affect the results on the upper bounds of $c$. Within the adopted GP estimation method of the upper bound on $c$, in order to be able to either reject or recover the status of EFT theories,  additional analysis is still required, starting by the consideration of other datasets and kernels with different priors for the hyperparameters. Our study, as reported here, is to be pondered as a first, albeit already revealing, attempt to show the benefits of using GP techniques in the study of Swampland criteria for the dark-energy dominated universe.

\section*{Acknowledgements}
We thank Maurice van Putten for interesting comments. EE has been partially supported by MINECO (Spain), Project FIS2016-76363-P, by the Catalan Government 2017-SGR-247, and by the CPAN Consolider Ingenio 2010 Project. MK is supported in part by a Chinese Academy of Sciences President's International Fellowship Initiative Grant (No. 2018PM0054).

\end{document}